# ELECTROMECHANICAL COUPLING COEFFICIENT: NEW APPROACH TO STUDY AUXETIC PIEZOELECTRIC HARVESTERS

*Grégoire Forges[1,2], David Gibus[1], Adrien Morel[1], Adrien Badel[1] and Hélène Debéda[2]*
[1]Université Savoie Mont-Blanc, SYMME, F-74000 Annecy, France and
[2]Université de Bordeaux, CNRS, Bordeaux INP, IMS, UMR 5218, 33400 Talence, France

## ABSTRACT

This work introduces a novel methodology to assess the performance of Piezoelectric Energy Harvesters (PEHs) in order to study auxetic enhancement possibilities. For this purpose, a new approach for evaluating the intrinsic effective Electromechanical Coupling Coefficient (EMCC) of piezoelectric layers is presented. As the current assessment methods are questioned under resonance exposures, theoretical models are presented to suggest what characteristics the harvested power will depend on. A two axis graph is introduced to enable the comparison of different PEHs. The method is finally applied to PEHs with different types of substrates: filled, hollow and auxetic. First results show that, generally, auxetic structures might not increase the intrinsic EMCC but only improve the elastic energy ratio in the piezoelectric layers.

## KEYWORDS

Energy Harvesting, Piezoelectricity, Auxetic Structures, Performance Assessment, Electromechanical Coupling Coefficient, Piezoelectric Coefficient.

## INTRODUCTION

The development of the Internet of Things (IoT) and wireless sensor nodes require energy harvesting from local sources due to the unrealistic battery maintenance. Ambient vibration energy harvesting may be achieved through electromechanical conversion using electromagnetic, electrostatic, and piezoelectric transducers. Piezoelectric Energy Harvesters (PEHs) are widely used, in particular those with mechanical resonators for high energy density. Lead-based ceramic materials, such as $Pb(Zr_xTi_{1-x})O_3$ (PZT), are the most commonly used. However, due to lead toxicity, it is essential to develop lead-free piezoelectric harvesters with comparable performance. Lead-free piezoelectric materials being feebly coupled, auxetic substrate structures with negative Poisson's ratio (figure 1), such as re-entrant honeycomb arrays, were studied for their low (under $-1$) Poisson's ratio [1].

Recent research on PEHs with auxetic substrates indicates improved harvested power performance compared to those with filled substrates. This is attributed to improved strain when stretching out of resonance (quasi-static) [2], [3]. Besides, the improvement with auxetic substrates could be explained by a better exploitation of radial mode EMCC ($k_p^2$) and/or a decrease in the substrate's stiffness [4]–[6]. Hence, with auxetic substrates having holes, the substrate's stiffness decreases. It remains however unclear whether the Electromechanical Coupling Coefficient (EMCC) improvement with auxetic substrates mainly comes from a decrease in stiffness or from a better exploitation of the piezoelectric material. In this work, we will first discuss the current methods used to study novel complex structures and identify their limitations. Then, a new approach based on finite elements simulation will be proposed to assess the performance. Using this method, we will compare the performance of PEHs with three different types of substrates: filled, hollow, and auxetic.

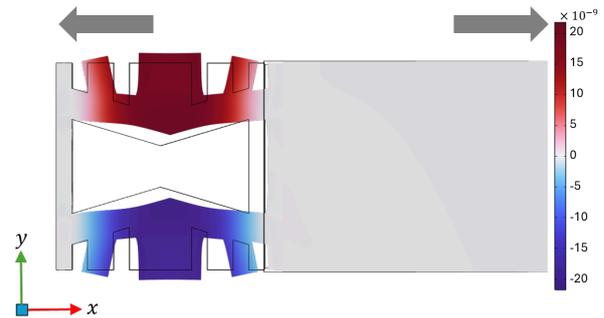

*Figure 1: Cantilever with auxetic design such as re-entrant honeycomb array. Grey arrows show axial stretch direction. Red and blue represent the displacement under $\vec{y}$ direction. Modal analysis simulation results in imprecise scale.*

## PERFORMANCE ASSESSMENT
**Literature discussion**

In the literature, the power improvement of PEHs with auxetic substrates is attributed to the additional strain in the lateral direction [4]. Consequently, the power delivery is proportional to the square of strain in both lateral and longitudinal directions as described in equation (1), $\sigma_{11}$ and $\sigma_{22}$ being respectively the strain in longitudinal and lateral directions [3]. Li *et al.* used quasi-static loading instead of resonance and it seems to be consistent.

$$P \propto (\sigma_{11} + \sigma_{22})^2 \quad (1)$$

However, at resonance, the power delivery is influenced by the mechanical quality factor ($Q_m$) and the global EMCC ($k_e^2$) [7] in a limit as shown on figure 2. The limit is moreover dependent on the mass ($M$) of the system, the acceleration ($\gamma$), the resonance pulsation ($\omega_0$) and $Q_m$, as described in equation (2).

$$P_{lim} = \frac{M\gamma^2 Q_m}{8\omega_0} \quad (2)$$

Chen *et al.* [4] considered two auxetic PEHs labelled A and B with resonance frequencies $f_{0,A} = 43.33$ Hz and $f_{0,B} = 36.48$ Hz. The authors found the maximum output power $P_{\max,A} = 4.34$ mW and $P_{\max,B} = 5.16$ mW while strain were $(\sigma_{11} + \sigma_{22})_A = 6.7$ MPa and $(\sigma_{11} + \sigma_{22})_B =$

8.42 MPa. Here, the power is not proportional to the square of strain but rather to the decrease in the resonance frequencies. More precisely, the powers are inversely proportional (equation (3)), which is consistent with equation (2).

$$\frac{P_{\max,B}}{P_{\max,A}} = \frac{f_{0,A}}{f_{0,B}} = 1.19 \quad (3)$$

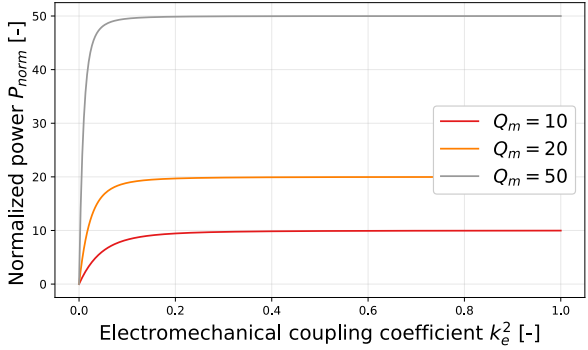

Figure 2: Normalized Power to the global EMCC for different values of $Q_m$ with an optimal purely resistive load. $P_{norm} = max(P)Q_m/P_{lim}$, with P the output power.

Hence, auxetic structures could increase output power of PEHs by increasing $k_e^2$ in a limit described by equation (2), regardless of the lateral and longitudinal stress distribution. However, impactful parameters remain unclear as equation (1) seems incomplete. In the next section, auxetic behavior's impact on the material EMCC ($k_{e,piezo}^2$) will be studied, as well as its influence on the elastic energy ratio in the piezoelectric elements.

**Impact of the strain and stress distribution on the material EMCC**

According to [8] and [9], the strain and stress distribution impact $k_e^2$ in two ways: firstly, on the equivalent material coefficients, and secondly, on the elastic energy. Different material coupling modes can be exploited for piezoelectric cantilevers, depending on the length ($l$) and the width ($w$) of the beam. $\boldsymbol{T}$ and $\boldsymbol{S}$ are the stress and strain vectors.

- With $l \gg w$, there is no lateral stress but lateral strain ie. $T_2 = 0$ and $S_2 \neq 0$. The plane stress transverse mode material EMCC $\left(k_{31}^2\right)^l$ is calculated in equation (4), with $d_{31}$ the piezoelectric constant, $s_{11}^E$ the mechanical stiffness at constant electric field and $\varepsilon_{33}^T$ the electrical permittivity at constant stress.

- With $w \gg l$, there is no lateral strain but lateral stress ie. $S_2 = 0$ and $T_2 \neq 0$. The plane strain alternative transverse mode material EMCC $\left(k_{e,31}^2\right)^w$ is calculated as in equation (5), with $e_{31}^{ef}$ the effective piezoelectric constant, $c_{11}^{ef}$ the effective compliance and $\varepsilon_{33}^{ef}$ the effective permittivity.

$$\left(k_{31}^2\right)^l = \frac{d_{31}^2}{s_{11}^E \varepsilon_{33}^T} \quad (4) \qquad \left(k_{e,31}^2\right)^w = \frac{\left(e_{31}^{ef}\right)^2}{c_{11}^{ef} \varepsilon_{33}^{ef}} \quad (5)$$

- If there is lateral strain, lateral stress, $S_1 = S_2$ and $T_1 = T_2$, the radial extensional mode EMCC ($k_p^2$) is approached [9]. It can be expressed as equation (6), with $\nu = -s_{12}^E/s_{11}^E$.

$$k_p^2 = \frac{2}{1-\nu}\frac{d_{31}^2}{s_{11}^E \varepsilon_{33}^T} \quad (6)$$

The equivalent material coefficients are considered for perfect modes and $\left(k_{31}^2\right)^l < \left(k_{31}^2\right)^w < k_p^2$ (see figure 4). If piezoelectric transducers are integrated on a substrate, this latter's influence must be taken in consideration. In particular, auxetic substrates will influence the strain in the piezoelectric elements as the negative Poisson's ratio will lead the piezoelectric material mode to approach the radial mode ($k_{e,piezo}^2$ will therefore tend to $k_p^2$). In order to evaluate $k_{e,piezo}^2$, the next section presents finite elements simulations method, beginning from an analytical point of view.

**Method to determine the material EMCC**

In [8], $k_{e,piezo}^2$ and $k_e^2$ are respectively given by equations (7) and (8), where $\alpha$ is the coupling term, $K_{piezo}$ the piezoelectric elements' equivalent stiffness, $K$ the global stiffness, and $C_p$ the capacitance.

$$k_{e,piezo}^2 = \frac{\alpha^2}{K_{piezo}C_p} \quad (7) \qquad k_e^2 = \frac{\alpha^2}{KC_p} \quad (8)$$

According to [10], $K$, with $K = K_{piezo} + K_{sub}$ ($K_{sub}$, the substrate's stiffness) is deduced from the global short circuit elastic energy ($U$) in equation (9), with $V_p$ the piezoelectric elements' volume, $V_s$ the substrate's volumes, $\boldsymbol{S^t}$ the transposed strain vector, $\boldsymbol{T}$ the stress vector, and $R(t)$ the generalized mechanical coordinate. Equation (8) can be written as equation (10).

$$U = \underbrace{\frac{1}{2}\int_{V_p} \boldsymbol{S^t T} dV_p}_{U_{piezo}=\frac{1}{2}K_{piezo}R(t)^2} + \underbrace{\frac{1}{2}\int_{V_s} \boldsymbol{S^t T} dV_s}_{K_{sub}R(t)^2} \quad (9)$$

$$k_e^2 = k_{e,piezo}^2 \frac{K_{piezo}}{K_{piezo} + K_{sub}} \quad (10)$$

In order to determine $k_{e,piezo}^2$ with COMSOL Multiphysics, the stiffness terms can be transformed into elastic energies leading to equation (11), where $U_{piezo}$ is the short circuit elastic energy in the piezoelectric elements.

$$k_e^2 = k_{e,piezo}^2 \frac{U_{piezo}}{U} \quad (11)$$

The elastic energy ratio can be computed by modal analysis in finite elements simulation, as well as $k_e^2$ with open and short circuit frequencies ($f_{oc}$ and $f_{sc}$), using equation (12), which leads to $k_{e,piezo}^2$ using equation (11).

$$k_e^2 = \frac{f_{oc}^2 - f_{sc}^2}{f_{sc}^2} \quad (12)$$

According to the model, $k_{e,piezo}^2$ and the elastic energy ratio are two important parameters. On one hand, by adding lateral displacement, the improvement could be attributed to approaching radial mode exposure. On the other hand, the elastic energy ratio could be increased by making holes in the substrate (reducing $K_{sub}$) or by increasing strain in the piezoelectric elements (increasing $K_{piezo}$). The objective is to clarify whether the performance is attributable to better coupling, substrate's lower stiffness, improved strain and stress, or a combination of these factors. To complete the reflection, the method is applied to a bimorph wide cantilever to observe how an elastic energy ratio drop influences the performance.

# MODEL IMPLEMENTATION ON FINITE ELEMENTS SIMULATIONS
## Application to a bimorph wide cantilever

A structure made with a wide filled substrate was designed to study lead-free piezoelectric harvesters $(K_{0.5}Na_{0.5})NbO_3$ (KNN). It was observed by [8], that with a gap between the clamping part and the piezoelectric elements (see figure 3), $k_e^2$ would significantly drop. The method enables to determine the two parameters and attempts to explain the cause in the drop of $k_e^2$.

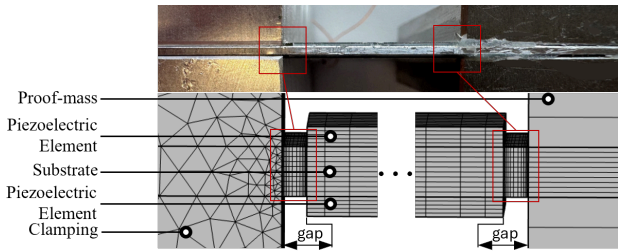

*Figure 3: Prototype and finite elements simulation's mesh on a bimorph cantilever PEHs with large clamp/beam and mass/beam gaps of 0.5 mm. Computed by finite elements simulation on COMSOL Multiphysics.*

A two axis graph is introduced on figure 4 to compare PEHs. The results drawn on the graph suggest that although $k_e^2$ is influenced by the energy concentrated in the gap, as the elastic energy ratio decreases, the piezoelectric material's mode remains the same.

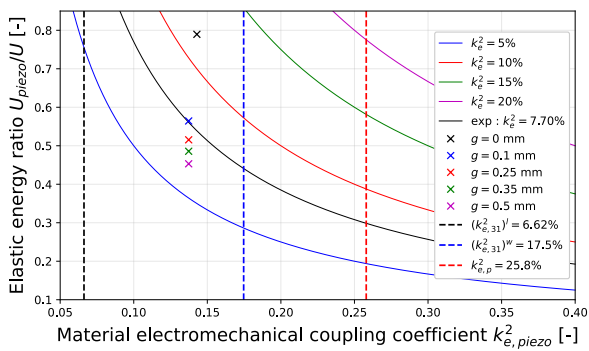

*Figure 4: Influence of the gap on the elastic energy ratio and $k_{e,piezo}^2$. Isolines represent values of $k_e^2$, the dotted lines represent the ideal $k_{e,piezo}^2$ for long/wide transverse and radial modes. The dots are the values for the different gaps and are computed by COMSOL Multiphysics. The experimental value of $k_e^2$ is represented by the black isoline.*

Thus, $k_e^2$ can significantly evolve without any change on $k_{e,piezo}^2$, since it only depends on the elastic energy ratio. Futhermore, different types of substrates can be compared: filled, hollow and auxetic.

## Comparison of different substrates, results and discussion

Figure 1 introduced the concept of a negative Poisson's ratio by showing the lateral displacement of the auxetic substrate under axial stretch. However, it is important to note that after adding the piezoelectric elements, the material's stiffness significantly prevents the substrate from influencing the strain in the piezoelectric elements. Therefore, it is crucial to compute the displacement with and without piezoelectric elements, to confirm the auxetic behavior.

In figure 5, two different substrates are observed. Both have the exact same filled area. On the left, a hollow plate and on the right, the substrate with auxetic structure from figure 1. The objective of the comparison is to determine how the auxetic pattern influences the defined parameters.

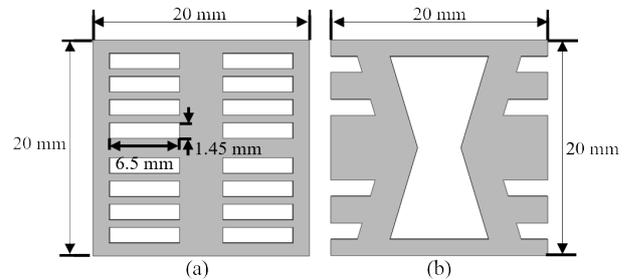

*Figure 5: Illustrations of the compared substrates having both a filled area of 249.35 $mm^2$ which represents 62.34 % of the square.*

On figure 6 and figure 7, displacement field on $\vec{y}$ axis is computed under axial resonance mode and under bending resonance mode.

Firstly, in figure 6, a reversed lateral displacement can be observed on the auxetic beam. However, it is not perfect as some compression remains.

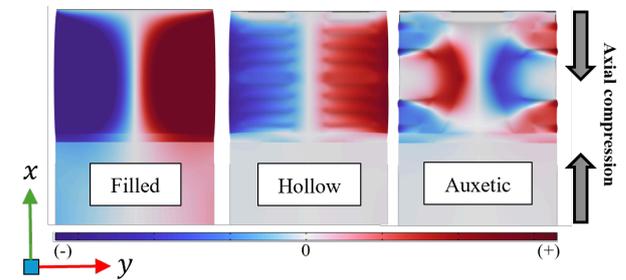

*Figure 6: Displacement in the piezoelectric beams under axial resonance mode. From left to right: filled, hollow and auxetic substrates.*

Secondly, in figure 7 under bending resonance mode, even if less displacement is observed for the auxetic substrate, it seems that auxetic behavior cannot be observed on the piezoelectric elements.

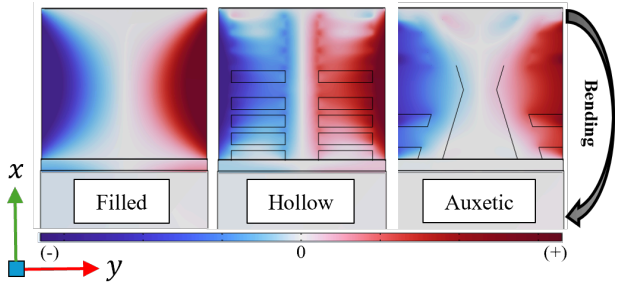

*Figure 7: Displacement in the piezoelectric beams under bending resonance mode. From left to right: filled, hollow and auxetic substrates.*

The computed results are given in table 1 and table 2. Under axial mode, although $k_{e,piezo}^2$ does not rise for the auxetic structure, the elastic energy ratio is multiplied by 5. Thus, $k_e^2$ increases proportionally. Otherwise, under bending mode, even if the elastic energy ratio slightly increases as the material EMCC decreases, the global EMCC drops compared to the hollow beam.

*Table 1: Comparison of the simulated performance for the three kinds of substrates under axial resonance mode.*

| Substrate types : | Filled | Hollow | Auxetic |
|---|---|---|---|
| $f_{oc}$ (Hz) | 10670 | 7103 | 6922 |
| $f_{sc}$ (Hz) | 10641 | 7049 | 6838 |
| $k_e^2$ | 0.55 % | 1.54 % | 2.45 % |
| Energy Ratio | 5.13 % | 16.73 % | 24.96 % |
| $k_{e,piezo}^2$ | 10.72 % | 9.21 % | 9.82 % |

*Table 2: Comparison of the simulated performance for the three kinds of substrates under bending resonance mode.*

| Substrate types : | Filled | Hollow | Auxetic |
|---|---|---|---|
| $f_{oc}$ (Hz) | 111.78 | 81.57 | 81.9 |
| $f_{sc}$ (Hz) | 110.47 | 79.60 | 80.23 |
| $k_e^2$ | 2.38 % | 5.02 % | 4.32 % |
| Energy Ratio | 18.00 % | 40.62 % | 42.09 % |
| $k_{e,pieo}^2$ | 13.20 % | 12.36 % | 10.26 % |

These results suggest that the designed auxetic structure has an impact on the PEHs performance. Though, this improvement is attributed to the rise of energy concentration in the piezoelectric elements and not to a better exploitation of the piezoelectric material.

## CONCLUSION

A novel approach was introduced to assess the performance of PEHs with auxetic substrates. After testing the method using finite elements simulation software on a wide cantilever, filled, hollow, and auxetic substrates were compared. It was highlighted that, even with auxetic substrate, it is hard to reverse strain in the piezoelectric elements. Finally, the designed auxetic substrate did not permit to improve piezoelectric exposure in the material. Such a method should be employed to assess the performance of auxetic structures in order to find out their impact on piezoelectric performance.

## ACKNOWLEDGEMENTS

This work was supported by the French National Agency for Research under grant ANR-22-CE51-0030.

## CONTACT

Grégoire Forges, tel: +33 (0)4 50 09 65 67;
gregoire.forges@univ-smb.fr